\documentclass[aps,prl,prabib,twocolumn,showpacs]{revtex4}
\usepackage{amsmath,verbatim}
\usepackage{amssymb}
\usepackage{amsfonts}
\begin{document}

\title{Virial theorems for trapped cold atoms}

\author{F\'{e}lix Werner}
\affiliation{Laboratoire Kastler Brossel, \'{E}cole Normale Sup\'{e}rieure, Universit\'{e} Pierre et Marie Curie-Paris 6, CNRS,
 24 rue Lhomond, 75231 Paris Cedex 05, France}

\date{\today}
\begin{abstract}
We present a general virial theorem for quantum particles
with arbitrary zero-range or finite-range interactions
in an arbitrary external potential.
We deduce virial theorems for several situations
relevant to trapped cold atoms:
zero-range interactions with and without Efimov effect, hard spheres,
narrow Feshbach resonances, and finite-range interactions. 
If the scattering length $a$ is varied adiabatically in the BEC-BCS crossover,
we find that the trapping potential energy as a function of $1/a$ has an inflexion point
at unitarity.

\end{abstract}


\maketitle



In quantum mechanics, 
zero-range interactions
can be expressed as boundary conditions on the many-body wavefunction in the limit of vanishing interparticle distance~\cite{Delta}.
These boundary conditions define the {\it domain} of the Hamiltonian,
i.~e. the set of wavefunctions on which the Hamiltonian is allowed to act.
The Hamiltonian of a zero-range model differs from the non-interacting Hamiltonian only by its domain.
In $3D$,
the zero-range model has a long history in
nuclear physics going back to the work of
Wigner, Bethe and Peierls 
on the $2$-nucleon problem~\cite{WignerBethePeierls}.

Zero-range interactions provide an accurate description of cold atom experiments~\cite{Varenna,YvanVarenna,GoraVarenna}.
In particular, two-component fermionic atoms in $3D$ at a broad Feshbach resonance
are well described  by zero-range interactions of scattering length $a=\infty$.
This so-called unitary limit is completely universal,
e.~g. the superfluid transition temperature is a universal number times the Fermi energy~\cite{ThomasCapa,Tc,thomas_entropie}.

A new ingredient in cold atomic systems 
with respect to nuclear physics
is the external trapping potential.
For the unitary Fermi gas in a harmonic trap,
the virial theorem
\begin{equation}
E = 2 E_{\rm tr}
\label{1}
\end{equation}
was recently shown theoretically
and experimentally~\cite{Chevy,YvanWeb,WernerPRA,thomas_virial}.
Here $E$ is the total energy
and $E_{\rm tr}$ is the trapping potential energy.

On the other hand, the traditional virial theorem
does not concern zero-range interactions, but
 more usual interactions described by a potential energy $U(\vec{r}_1,\dots,\vec{r}_N)$, where the domain is simply a set of smooth functions.
It states that the kinetic energy $T$ is one half of the virial:
\begin{equation}
\left< T  \right> = \frac{1}{2}\left< \sum_{i=1}^N {\bf r_i} \cdot {\bf \nabla_{r_i}}U\right>
\label{virial_V}
\end{equation}
for any eigenstate; implying 
$\left< T  \right> = n/2\,\left< U \right>$
if $U$ is a homogeneous function of degree $n$.
This theorem is as old as many-particle quantum mechanics~\cite{heisenberg},
and is used e.~g. to  understand the properties of many-electron atoms~\cite{carbone}.

In this paper, we present a general virial theorem for 
a Hamiltonian with an arbitrary domain.
In
the particular case where the domain does not depend on any length scale,
we recover the virial theorem for the unitary gas Eq.~(\ref{1})
and the traditional virial theorem Eq.~(\ref{virial_V}).
By considering the case of a more general domain,
we find  new virial theorems  for several interactions relevant to cold atoms:
zero-range interactions of arbitrary scattering length with or without Efimov effect,
hard spheres, 
narrow Feshbach resonances,
and finite-range interactions.
Our theorems hold for any trapping potential, in any space dimension.
They are valid not only for each eigenstate, but also at thermal equilibrium provided the entropy $S$ is kept constant.
For zero-range interactions without Efimov effect, the virial theorem implies that for any $S$, the function $E_{\rm tr}(1/a,S)$
has an inflexion point at the unitary limit $1/a=0$.

\noindent \underline{\bf General Virial Theorem.} Let us consider a  quantum problem of $N$ particles, with arbitrary statistics and dispersion relations.
The position ${\bf r_i}$ of particle $i$
is a vector of arbitrary dimension, with continuous or discrete coordinates.
We consider a general Hamiltonian
\begin{equation}
H=H'+U({\bf r_1},\dots,{\bf r_N})
\label{eq:H=}
\end{equation}
where
\\$\bullet$ $H'$ and its domain depend on $p$ parameters 
$l_1,\ldots,l_p$ which have the dimension of a length,
on $\hbar$, and on some arbitrary fixed mass $m$
\\$\bullet$ $U({\bf r_1},\dots,{\bf r_N})$ is an arbitrary function,
which is sufficiently 
regular
 so that
the domains of $H$ and $H'$ coincide.
\\Then, as shown below:
 \begin{equation}
E = 
\left< U+\frac{1}{2}\sum_{i=1}^N {\bf r_i}\cdot {\bf \nabla_{\bf r_i}}U \right>  - \frac{1}{2}\sum_{q=1}^p l_q \frac{\partial E}{\partial l_q}
\label{thm}
\end{equation}
for any stationary state of energy $E$,
the partial derivatives $\partial E/\partial l_q$ being taken for a fixed function $U$.

To derive the above theorem, we use dimensional analysis to rewrite $U$ as
\begin{equation}
U({\bf r_1},\dots,{\bf r_N})= \frac{\hbar^2\lambda^2}{m} f(\lambda{\bf r_1},\dots,\lambda{\bf r_N})
\label{eq:U}
\end{equation}
where $\lambda$ has the dimension of the inverse of a length, and $f$ is dimensionless function.
The theorem then follows from the following two relations:
\begin{eqnarray}
\lambda \frac{\partial E}{\partial \lambda} &=& \left<  2 U+\sum_{i=1}^N {\bf r_i}\cdot {\bf \nabla_{\bf r_i}}U \right>
\label{lemme1}
\\
  \lambda \frac{\partial E}{\partial \lambda} &=& 2 E + \sum_{q=1}^p l_q \frac{\partial E}{\partial l_q}.
\label{lemme2}
\end{eqnarray}
Here the partial derivatives with respect to $\lambda$ are taken for a fixed function $f$ and for fixed $l_1,\dots,l_p$.
\\Eq.~(\ref{lemme1})
follows from the Hellmann-Feynman theorem~\cite{Son} and from Eq.~(\ref{eq:U}).
 The Hellmann-Feynman theorem holds if
the derivative
$\partial \left|\psi\right>/\partial\lambda$
of the considered eigenstate
 belongs to the domain of $H$.
We expect this to be true in all situations considered in this paper.
\\Eq.~(\ref{lemme2})
follows from the fact that, by dimensional analysis, the energy writes
\begin{equation}
E(l_1,\ldots,l_p,[U] ) = \frac{\hbar^2\lambda^2}{m} F(\lambda l_1,\ldots,\lambda l_p,[f])
\end{equation}
where $F$ is a dimensionless functional.

The traditional virial theorem Eq.~(\ref{virial_V}) is recovered by applying the general virial theorem 
to the case where:
\\$\bullet$ The operator $H'$ in Eq.~(\ref{eq:H=}) reduces to the kinetic energy
\begin{equation}
T=-\sum_{i=1}^N \frac{\hbar^2}{2 m_i} \Delta_{\bf r_i},
\end{equation}
$m_i$ being the mass of particle $i$;
\\$\bullet$ The domain is simply a set a wavefunctions which are smooth when particles approach each other.
\\Since this domain does not depend on any length scale, the second term on the right-hand-side of Eq.~(\ref{thm}) vanishes, and the result Eq.~(\ref{virial_V}) follows.

\noindent \underline{\bf Virial theorems for trapped cold atoms.}
In what follows we restrict to the experimentally relevant case where $U$ is a sum of trapping potentials:
\begin{equation}
U({\bf r_1},\dots,{\bf r_N})=\sum_{i=1}^N  U_i({\bf r_i}),
\end{equation}
and we rewrite the general virial theorem Eq.~(\ref{thm}) as:
\begin{equation}
E = 2 \tilde{E}_{\rm tr} - \frac{1}{2} \sum_{i=1}^p l_i \frac{\partial E}{\partial l_i},
\label{thm2}
\end{equation}
where
\begin{equation}
\tilde{E}_{\rm tr} \equiv \frac{1}{2}
\sum_{i=1}^N
 \left< 
  U_i({\bf r_i})+\frac{1}{2}{\bf r_i}\cdot {\bf \nabla}U_i({\bf r_i})
\right>.\label{Etrap}
\end{equation}
If each $U_i$ is a harmonic trap,
then $\tilde{E}_{\rm tr}$ reduces to the trapping potential energy:
$\tilde{E}_{\rm tr} = 
\sum_{i=1}^N \left< U_i({\bf r_i}) 
\right>= E_{\rm tr}$.


\noindent {\bf A. Zero-range interactions.}
We now assume that each pair of particles either interacts {\it via} a zero-range interaction of scattering length $a$, or does not interact.
Zero-range interactions are well-known in $1D$~\cite{LiebLiniger,Gaudin},
$2D$~\cite{LudoMaxim2D}
and $3D$~\cite{WignerBethePeierls,Efimov,Albeverio,Petrov4corps,Pethick,WernerPRA,WernerPRL}.

\noindent{\bf A.1 Universal states.}
We call {\it universal state}
a stationary state of the zero-range model
which depends only on the scattering length.
All eigenstates are believed to be universal in $1D$ and $2D$
(\cite{LiebLiniger,Gaudin,Leyronas2D} and references therein)
and in $3D$ for fermions with two components of equal masses~\cite{Efimov,thomas_virial,Chevy,YvanWeb,Varenna,Albeverio,YvanVarenna,GoraVarenna,WernerPRA,WernerPRL,Blume,Stecher,Petrov4corps,pandharipande,giorgini,Juillet,Tc,YvanBoite,GrimmCrossover,SalomonCrossover,ThomasCapa,HuletPolarScience,JinPotentialEnergy,GrimmVarenna,ENSvarenna,thomas_entropie}
or unequal masses with a mass ratio not too far from one~\cite{GoraVarenna,Stecher}.
For $3$ bosons in $3D$
there are both non-universal efimovian states and universal states~\cite{Pethick,WernerPRL}.

In the Hilbert space generated by universal states,
the domain of the Hamiltonian depends only on the scattering length. Thus
Eq.~(\ref{thm2}) gives for any universal state:
\begin{equation}
E = 2 \tilde{E}_{\rm tr} - \frac{1}{2} a \frac{\partial E}{\partial a},
\label{viriel_a1}
\end{equation}
or equivalently
\begin{equation}
E = 2 \tilde{E}_{\rm tr} + \frac{1}{2a} \frac{\partial E}{\partial (1/a)}.
\label{viriel_a}
\end{equation}
This result generalizes the virial theorem Eq.~(\ref{1})
to an arbitrary scattering length, trapping potential and space dimension.
Thus it also applies to quantum gases in low dimensions~(\cite{RevueJean,Weiss1dTonks,JeanNJP,krauth}
 and refs. therein).
For the case of $2$-component fermions in $3$ dimensions and power-law traps,
this result is also contained in two recently submitted works:
it was found independently by S.~Tan in \cite{TanViriel}
and rederived using a method similar to ours in \cite{BraatenViriel}.

For $a=\infty$ (which is the unitary limit in $3D$ and the non-interacting limit in $1D$ and $2D$),
Eq.~(\ref{viriel_a}) becomes:
\begin{equation}
E = 2 \tilde{E}_{\rm tr}.
\label{eq:yvan}
\end{equation}
This generalisation of Eq.~(\ref{1}) to an arbitrary trap was obtained by Y.~Castin~(unpublished), and
is also contained in the recent independent work of J.~Thomas~\cite{ThomasVirielTheorie}.
Of course it also holds for $a=0$ (which is the Tonks-Girardeau limit in $1D$ and the non-interacting limit $2D$ and $3D$) in accordance with Eq.~(\ref{viriel_a1}).

Taking 
the second derivative of Eq.~(\ref{viriel_a})
we obtain:
\begin{equation}
\left. \frac{\partial^2 \tilde{E}_{\rm tr}}{\partial(1/a)^2}\right|_{a=\infty}=0,
\label{eq:inflex}
\end{equation}
which means that generically the curve $\tilde{E}_{\rm tr}(1/a)$
has an inflexion point exactly at the unitary limit $1/a=0$.

We  can also rewrite 
Eq. (\ref{viriel_a}) in an integral form:
\begin{equation}
a_2^{\phantom{2}2} E(a_2) - a_1^{\phantom{2}2} E(a_1) =
-4 \int_{1/a_1}^{1/a_2}  a^3 \tilde{E}_{\rm tr}(a)\, d(1/a),
\label{eq:integral}
\end{equation}
which is likely to have a better signal-to-noise ratio than Eq.~(\ref{viriel_a})
when applied to experiments or numerics.

\noindent
{\bf A.2 Efimovian states.}
The boundary condition in the limit where two particles approach each other is called Bethe-Peierls boundary condition (BPbc).
For $3$ bosonic or distinguishable particles,
there exists Efimov bound states~\cite{Efimov},
and the domain of the zero-range model is defined not only by the BPbc in the limit where two particles approach each other, but also by an additional boundary condition in the limit where all three particles approach each other.
While the BPbc depends on the scattering length $a$,
this additional boundary condition depends on a $3$-body parameter
which we call $R_t$
and has the dimensions of a length~\cite{WernerPRL,WernerThese}.
The resulting $2$-parameter model is known to be self-adjoint and physically meaningful for $N=3$ particles
~\cite{Efimov,Pethick,WernerPRL,WernerThese,Albeverio}.
The case $N\geq4$ is still controversial~\cite{4corpsHammerPlatter}.

For this model,
the general virial theorem Eq.~(\ref{thm2}) gives:
\begin{equation}
E = 2 \tilde{E}_{\rm tr} + \frac{1}{2} \left[\frac{1}{a} \frac{\partial E}{\partial (1/a)} - R_t \frac{\partial E}{\partial R_t}\right].
\label{viriel_efi}
\end{equation}
For $a=\infty$ this reduces to
\begin{equation}
E = 2 \tilde{E}_{\rm tr}  - \frac{R_t}{2} \frac{\partial E}{\partial R_t}.
\label{viriel_Rt}
\end{equation}

We now apply this to the unitary $3$-boson problem in an isotropic harmonic trap, which is exactly solvable~\cite{WernerPRL,Pethick}. The spectrum is $E=E_{\rm CM}+{\cal E}\,\hbar\omega$ where $E_{\rm CM}$ is the energy of the center-of-mass and ${\cal E}$ solves:
\begin{equation}
{\rm arg}\, \Gamma\left( \frac{1+s-{\cal E}}{2} \right) = -|s|{\rm ln}\, R_t + {\rm arg}\,\Gamma(1+s)\ \ {\rm mod}\ \pi,
\end{equation}
$s\simeq i \cdot 1.00624$ being the only solution $s\in i
\cdot(0;+\infty)$ of the equation:
$s \cos\left(s \pi/2\right) - 8/\sqrt{3}\, \sin\left(s \pi/6 \right) = 0$.
This allows to calculate $\partial {\cal E}/\partial R_t$, and Eq.~(\ref{viriel_Rt}) then gives \cite{cond-mat}:
\begin{equation}
E_{\rm tr} = \frac{1}{2} \left( 
E+\frac{|s|}{{\rm Im}\, \psi\left(\frac{1+s-{\cal E}}{2}\right)}
\right)
\label{eq:Etrap}
\end{equation}
where $\psi$ is the digamma function.
But we can also express $E_{\rm tr}$ using the wavefunction, which has a simple expression in terms of the Whittaker $W$ function~\cite{WernerPRL}; the result agrees with Eq.~(\ref{eq:Etrap}) provided that~\cite{cond-mat}:
\begin{eqnarray}
\int_0^\infty dx\,\left[W_{\frac{\cal E}{2},\frac{s}{2}}(x)\right]^2 &=&
\left( {\cal E}\cdot {\rm Im}\,\psi\left(\frac{1+s-{\cal E}}{2}\right)+|s|\right)
\nonumber
\\
&  &\cdot\frac{2\ \pi}{{\rm sinh}(|s|\pi)\, \left|\Gamma\left(\frac{1+s-{\cal E}}{2}\right)\right|^2} 
 .
\end{eqnarray}
Numerical checks confirm this relation.

\noindent
{\bf B. Hard sphere interactions.}
Here the domain is defined by the condition that the wavefunction vanishes if any interparticle distance is smaller than $a$.
Applying the general virial theorem with a single length scale gives:
\begin{equation}
E = 2 \tilde{E}_{\rm tr} - \frac{1}{2} a \frac{\partial E}{\partial a}.
\label{viriel_weak}
\end{equation}
Again, it can be useful to rewrite Eq. (\ref{viriel_weak}) in an integral form:
\begin{equation}
E(a) = \frac{4}{a^2} \int_0^a \, a' \tilde{E}_{\rm tr}(a') \, da' .
\label{eq:integral_weak}
\end{equation}
Within the $3D$ Gross-Pitaevskii theory, $a\,\partial E/\partial a$ is the interaction energy,
so that Eq.~(\ref{viriel_weak}) agrees with the virial theorem of~\cite{Stringari}.

\noindent
{\bf C. Finite-range interactions.}
We now consider models with two parameters, the scattering length $a$
and a range $l$.
Popular examples are the square-well interaction potential~\cite{giorgini},
separable potentials~\cite{WernerPRL},
and
Hubbard-like lattice models
where the lattice spacing $l$ plays the role of the interaction range~\cite{Tc,Juillet,YvanBoite}.
For such $2$-parameter models the general virial theorem gives:
\begin{equation}
E=2\tilde{E}_{\rm tr}+\frac{1}{2} \left[\frac{1}{a}\frac{\partial E}{\partial (1/a)}
-l\frac{\partial E}{\partial l}\right],
\label{virial_l}
\end{equation}
and for $a=\infty$:
\begin{equation}
E=2\tilde{E}_{\rm tr}-\frac{l}{2} \frac{\partial E}{\partial l}.
\label{virial_l_unitaire}
\end{equation}
Setting $E_0=\lim_{l\to 0}E(l)$, Eq.~(\ref{virial_l_unitaire}) implies
\begin{equation}
E_0=3 E - 4 \tilde{E}_{\rm tr}+O(l^2), 
\end{equation}
which
can be used to compute numerically $E_0$.
This method is simpler than the usual one, where
one computes $E$ for several values of $l$
and extrapolates linearly to $l=0$~\cite{YvanBoite,WernerPRL,Stecher}.

\noindent {\bf D. Effective range model and narrow resonances.}
The effective range model has two parameters, the scattering length $a$ and the effective range $r_e$.
For $r_e<0$, the  model describes a narrow Feshbach resonance~\cite{PetrovBosons,YvanVarenna,BraatenModels,MattiaPrivate,mora3corps}.
For $r_e\to 0^-$, the model has a limit cycle described by the zero-range model of Sec.~{\bf A.2}, with $R_t=C\,r_e$, where the constant $C$ was obtained numerically~\cite{PetrovBosons} and analytically~\cite{mora3corps}.
The model is expected to be hermitian for a modified scalar product, for $2$ particles~\cite{LudoOndeL} and $3$ particles~\cite{MattiaPrivate}.
Thus the Hellmann-Feynman theorem can be used and the general virial theorem
 holds, implying:
\begin{equation}
E = 2 \tilde{E}_{\rm tr} + \frac{1}{2} \left[\frac{1}{a} \frac{\partial E}{\partial (1/a)} - r_e \frac{\partial E}{\partial r_e}\right].
\label{viriel_re}
\end{equation}
For $r_e>0$,
 the effective range model is well-defined if $r_e$ is treated perturbatively~\cite{WernerThese},
and Eq.~(\ref{viriel_re}) holds, in agreement with Eq.~(\ref{virial_l}).

\noindent \underline{\bf At finite temperature.}
We will show that the above results
remain true at finite temperature, provided one considers adiabatic transformations.
For concreteness we restrict to zero-range interactions in the universal case.
We consider that each
eigenstate $n$ has an occupation probability
$p_n$.
We set  $\overline{E}=\sum_n E_n\, p_n$ and 
$\overline{\tilde{E}_{\rm tr}}=\sum_n (\tilde{E}_{\rm tr})_n\, p_n $.

Let us first recall the reasoning of Tan~\cite{Tan_Momentum_dE,TanViriel}:
for a finite system, in the limit where $a$ is varied infinitely slowly,
the adiabatic theorem implies that the $p_n$'s remain constant, so that
\begin{equation}
\sum_n \frac{\partial E_n}{\partial (1/a)}\, p_n = \frac{\partial }{\partial (1/a)}\sum_n E_n\, p_n.
\label{eq:dE}
\end{equation}
Tan concludes that $E$ and $\tilde{E}_{\rm tr}$ can be replaced
by their average values $\overline{E}$ and 
$\overline{\tilde{E}_{\rm tr}}$ in the virial theorem
 Eq.~(\ref{viriel_a}).

Alternatively, let us assume 
 that the $p_n$'s are given by the canonical distribution $p_n\propto e^{-E_n(a)/(k_B T)}$,
where the temperature $T$ varies with $a$ in such a way that the entropy
$S=-k_B \sum_n p_n \ln p_n$ remains constant.
According to the principles of thermodynamics,
this assumption is a good effective description of
adiabatic sweep
experiments where
 $a$ is changed at a rate much smaller than thermalisation rates and much larger than heating and evaporation rates~\cite{Carr,JinPotentialEnergy,thomas_entropie,GrimmVarenna,caveat_polkov}.
Under this assumption Eq.~(\ref{eq:dE}) also holds~\cite{Bogo}.
Thus:
\begin{equation}
\overline{E} = 2\, \overline{\tilde{E}_{\rm tr}} + \frac{1}{2a} 
\left(\frac{\partial \overline{E}}{\partial (1/a)}\right)_S\,.
\label{viriel_a_T}
\end{equation}
This result is physically consistent with Tan's conclusion.
Moreover it implies:
\begin{equation}
 \frac{\partial^2 \overline{\tilde{E}_{\rm tr}}}{\partial(1/a)^2}
\left(\frac{1}{a}=0,S\right)
=0,
\label{eq:inflex_T}
\end{equation}
\begin{equation}
a_2^{\phantom{2}2}\, \overline{E}(a_2,S) - a_1^{\phantom{2}2}\, \overline{E}(a_1,S) =
-4 \int_{1/a_1}^{1/a_2}  a^3 \overline{\tilde{E}_{\rm tr}}(a,S)\, d(1/a).
\label{eq:integral_T}
\end{equation}

\noindent \underline{\bf Experimental considerations.}
Both $E$ and $\tilde{E}_{\rm tr}$ are measurable.
Indeed, $\tilde{E}_{\rm tr}$ and
the trapping potential energy
$E_{\rm tr}$ can be deduced from an in-situ image of the density profile~\cite{GrimmCrossover,JinPotentialEnergy,HuletPolarScience,KetterleInSitu},
and
the released energy $E-E_{\rm tr}$
from a time of flight image
\cite{Thomas2002,SalomonCrossover,Weiss1dTonks,note_expansion}. 
By measuring $\overline{E}$ and $\overline{\tilde{E}_{\rm tr}}$,
and using
the virial theorem Eq.~(\ref{viriel_a_T}),
one could deduce
the quantity $(\partial \overline{E}/\partial (1/a))_S$~\cite{caveat}.
This quantity is also
related to the large-momentum tail of the momentum distribution~\cite{Tan_Momentum_dE}
and to the total energy~\cite{TanEnergetics}.

Moreover, Eqs.~(\ref{viriel_a_T},\ref{eq:inflex_T},\ref{eq:integral_T}) can be directly checked
by measuring $E(a)$ and $\tilde{E}_{\rm tr}(a)$ in an adiabatic sweep experiment.


I am grateful to S.~Tan and J.~Thomas for drawing my attention to~\cite{TanViriel,ThomasVirielTheorie,BraatenViriel},  and to
M.~Antezza,
S.~Biermann,
E.~Braaten,
Y.~Castin,
M.~Cheneau,
F.~Chevy,
J.~Dalibard,
B.~Derrida,
W.~Krauth,
F.~Lalo\"{e},
S.~Nascimb\`{e}ne,
M.~Olshanii,
A.~Ridinger,
B.~Roulet,
C.~Salomon,
R.~Sheshko,
S.~Tan,
L.~Tarruell, 
and
J.~Thomas
for discussions and comments.
LKB is a {\it Unit\'{e} Mixte de Recherche} of ENS, Universit\'{e}
 Paris 6 and CNRS.
Our research group is a member of IFRAF.


\begin{thebibliography}{30}
\bibitem{Delta}
Zero-range interactions can only be represented  by an interaction potential proportional to the Dirac distribution in $1D$, or  in perturbative approaches in $2D$ and $3D$.
\bibitem{note_expansion} If the expansion dynamics is known theoretically, then other experimental methods become available: energy can be precisely added to the gas~\cite{ThomasCapa,thomas_virial}, and $E_{\rm tr}$ and $\tilde{E}_{\rm tr}$ can be deduced from a time-of-flight image~\cite{thomas_virial,thomas_entropie}.
The expansion dynamics is known if hydrodynamics~\cite{ThomasCapa,thomas_virial}
or exact scaling solutions~\cite{Yvan} are applicable.
\bibitem{caveat} However this method breaks down at unitarity,
where~$E~-2\tilde{E}_{\rm tr}\to0$.
\end{thebibliography}
\end{document}